\documentclass[10pt,a4paper,reqno]{amsart}
\usepackage{amsthm}
\usepackage{amsmath}
\usepackage{amsmath, amssymb}
\usepackage{type1cm}
\usepackage{cite}
\usepackage{bm}
\usepackage{cases}
\theoremstyle{definition}
\newtheorem{defi}{Definition}[section]
\newtheorem{prop}{Proposition}[section]

\newtheorem{theorem}{Theorem}[section]
\newtheorem{corollary}{Corollary}[section]
\newcommand{\eqdef}{\stackrel{\mathrm{def}}{=}}
\begin{document} 
\title{Sum decomposition of divergence into three divergences}
\author{Tomohiro Nishiyama}
\begin{abstract}
Divergence functions play a key role as to measure the discrepancy between two points in the field of machine learning, statistics and signal processing. Well-known divergences are the Bregman divergences, the Jensen divergences and the f-divergences.
In this paper, we show that the symmetric Bregman divergence can be decomposed into the sum of two types of Jensen divergences and the Bregman divergence.
Furthermore, applying this result, we show another sum decomposition of divergence is possible which includes f-divergences explicitly.
\\

\end{abstract}
\date{}
\maketitle
\bibliographystyle{plain}
\section{Introduction}
Divergences are functions which measure the discrepancy between two points and play a key role in the field of machine learning, statistics, signal processing and so on.
Given a set $\Omega$ and $P,Q\in \Omega$, a divergence is defined as a function $D:\Omega\times \Omega\rightarrow \mathbb{R}$ which satisfies the following properties.\\
1. $D(P,Q)\geq 0 $ for all $P,Q\in  \Omega$.\\
2. $D(P,Q)=0\iff P=Q$.\\

In this paper, we mainly discuss about the following divergences.
\begin{itemize}
\item Bregman divergence\cite{bregman1967relaxation}. \\
$B_{F}(P,Q)\eqdef \sum_iF(p_i)-\sum_iF(q_i)-\sum_i F'(q_i)(p_i-q_i)$.
\item Symmetric Bregman divergence\cite{nielsen2009sided}. \\
$B_{F,\mathrm{sym}}(P,Q)\eqdef B_F(P,Q)+B_F(Q,P)$.
\item Jensen divergence\cite{nielsen2011burbea, burbea1980convexity}. \\
$J_{F,\alpha}\eqdef \alpha \sum_i F(p_i)+(1-\alpha)\sum_iF(q_i) -\sum_i F\bigl(\alpha p_i+(1-\alpha)q_i)$. 
\item $f$-divergence \cite{csiszar1967information,ali1966general}. \\
$D_f(P\| Q)\eqdef \sum_i q_i f(\frac{p_i}{q_i})$,
\end{itemize}
where $F(x)$ and $f(x)$ are a strictly convex functions, $f(1)=0$ and $\alpha\in(0,1)$.  

The main purpose of this paper is to show that some kind of divergences can be decomposed into three divergences.

In section 2, we introduce the 1-dimensional Bregman, the 1-dimensional symmetric Bregman, the multivariate Jensen and the multivariate convex conjugate Jensen divergences. 
Then, we show the sum of the 1-dimensional symmetric Bregman divergences is decomposed into the sum of the the multivariate Jensen, the multivariate convex conjugate Jensen and the Bregman divergences.

In section 3, we show the symmetric Bregman divergence is decomposed into the Jensen, convex conjugate Jensen and the Bregman divergences.
From this result, we can derive the following equation.
\begin{align}
\frac{1}{4}\mathrm{J}(P,Q)=\mathrm{JS}(P,Q)+\mathrm{H}^2(P,Q) + \mathrm{KL}\bigl(\frac{1}{2}(P+Q)\| \sqrt{PQ}\bigr),
\end{align}
where $\mathrm{J}(P,Q)$ denotes the Jeffrey's J-divergence\cite{jeffreys1946invariant}, $\mathrm{JS}(P,Q)$ denotes the Jensen-Shannon divergence\cite{lin1991divergence}, $\mathrm{H}^2(P,Q)$ denotes the squared Hellinger distance and $\mathrm{KL}(P\|Q)$ denotes the Kullback-Leibler divergence\cite{kullback1997information}.
We discuss about the details of this equation in section 3.

In section 4, we show another sum decomposition of divergence is possible.
This decomposition equation does not include the symmetric Bregman divergences explicitly but includes f-divergences explicitly.
From this result, we can derive the following equation for  $\sum_i p_i=\sum_i q_i=1$.
\begin{align}
\chi_N^2(P\|Q)=\mathrm{KL}(Q\| P)+\mathrm{IS}\biggl(1\|\frac{1}{\chi_N^2(P\|Q)+1}\biggr)+\biggl(\log\bigl(\chi_N^2(P\|Q)+1\bigr)-\mathrm{KL}(Q\| P)\biggr),
\end{align}
where $\chi_N^2(P\|Q)$ denotes Neyman chi-square divergence\cite{cichocki2010families} and $\mathrm{IS}(p\|q)$ denotes Itakura-Saito divergence\cite{itakura1968analysis}.
The last term $\biggl(\log\bigl(\chi_N^2(P\|Q)+1\bigr)-\mathrm{KL}(Q\| P)\biggr)$ is also a divergence function.
We discuss about the details of this equation in section 4.

Theorems proved in section 3 and 4 are main results of this paper.

\section{Preparation for proof}
\label{sec_base_sum_decomp}
\textbf{Notation of this section}
\begin{itemize}
\item Let $F:\mathrm{dom} F\rightarrow\mathbb{R}$ be differentiable and strictly convex functions and let $\mathrm{dom}F\subseteq\mathbb{R}$.
\item Let $\nu\in\{1,2,\cdots,N\}$ and $p_\nu,p,q\in\mathrm{dom}F$.
\item Let $\boldsymbol{\alpha}=(\alpha_1, \alpha_2, \cdots, \alpha_N)$ and $\bm{p}=(p_1,p_2,\cdots, p_N)$.
\item Let $\alpha_\nu> 0(\nu=1,2,3\cdots,N)$ be parameters which satisfy $\sum_\nu\alpha_\nu=1$.
\end{itemize}

First, we define the convex conjugate. 

\begin{defi}
The convex conjugate $F^\ast :\mathrm{dom} F^\ast\rightarrow\mathbb{R}$ is defined in terms of the supremum by
\begin{align}
F^\ast(x^\ast)\eqdef \sup_x \{\langle x^\ast, x\rangle - F(x)\},
\end{align}
where $\langle\cdot, \cdot\rangle$ denotes an inner product.
$F^\ast$ is a convex function.
\end{defi}
Because the derivative of $F$ is the maximizing argument, we get
\begin{align}
\label{convex conjugate}
x^\ast=\nabla F(x) \\ \nonumber
F(x)+F^\ast(x^\ast)=\langle x^\ast, x\rangle.
\end{align}

In the following, $x^\ast$ denotes $F'(x)$.

Then, we introduce the 1-dimensional Bregman and the 1-dimensional symmetric Bregman divergences.
\begin{defi}
\label{def_1dim_Bregman}
\begin{align}
B_F(p,q)\eqdef F(p)-F(q)-F'(q)(p-q)\\
\label{eq_def_1dim_sym_Bregman}
B_{F,\mathrm{sym}}(p,q)\eqdef B_F(p,q)+B_F(q,p)=(F'(p)-F'(q))(p-q)
\end{align}
\end{defi}

The 1-dimensional Bregman and the 1-dimensional symmetric Bregman divergences can be represented as follows by using convex conjugate\cite{amari2010information}.
\begin{prop}
\begin{align}
\label{Bregman_conjugate}
B_F(p,q)&=F(p)-F(q)-q^\ast(p-q)=F(p)+F^\ast(q^\ast)-q^\ast p \\
\label{eq_sym_Bregman_conjugate}
B_{F,\mathrm{sym}}&=(p^\ast-q^\ast)(p-q)
\end{align}
\end{prop}
These equations are easily proved by using (\ref{convex conjugate}).

\begin{prop}
\label{prop_reverse_Bregman}
\begin{align}
B_{F^\ast}(p^\ast,q^\ast)=B_F(q,p)
\end{align}
\end{prop}
\noindent\textbf{Proof.}
Combining $x=(F^\ast)'(F'(x))$ and  (\ref{Bregman_conjugate}), we get 
\begin{align}
B_{F^\ast}(p^\ast,q^\ast)&=F^\ast(p^\ast)-F^\ast(q^\ast)-q(p^\ast-q^\ast)\\ \nonumber
&=F^\ast(p^\ast)+F(q)-q^\ast q-q(p^\ast-q^\ast)=F(q)+F^\ast(p^\ast)-p^\ast q=B_F(q,p).
\end{align}

Then, we define the multivariate Jensen divergence.
\begin{defi}
\label{def_multi_Jensen}
\begin{align}
\label{eq_def_multi_Jensen}
J_{F,\boldsymbol{\alpha}}(\bm{p})\eqdef\sum_\nu \alpha_\nu F(p_\nu)-F(c)\\
c\eqdef  \sum_\nu \alpha_\nu p_\nu
\end{align}
\end{defi}
We show this function has divergence properties later.

The Jensen divergences can be represented as follows by using convex conjugate.
\begin{prop}
\begin{align}
\label{eq_Jensen_conjugate}
J_{F,\boldsymbol{\alpha}}(\bm{p})=\sum_\nu \alpha_\nu F(p_\nu)+F^\ast(c^\ast)-c^\ast c,
\end{align}
where $c\eqdef \sum_\nu \alpha_\nu p_\nu$.
\end{prop}
This equation is easily proved by using (\ref{convex conjugate}).

\begin{prop}
\label{prop_Jensen_Bregman}
The multivariate Jensen divergence is equal to the sum of the 1-dimensional Bregman divergences\cite{nielsen2009sided}.
\begin{align}
J_{F,\boldsymbol{\alpha}}(\bm{p})=\sum_\nu \alpha_\nu B_F(p_\nu, c)
\end{align}

\end{prop}
\noindent\textbf{Proof.}
By using  $\sum_\nu\alpha_\nu=1$ and (\ref{Bregman_conjugate}), we get
\begin{align}
\sum_\nu  \alpha_\nu B_F(p_\nu,c)=\sum_\nu \alpha_\nu F(p_\nu)+F^\ast(c^\ast)-c^\ast \sum_\nu  \alpha_\nu p_\nu \\ \nonumber
= \sum_\nu \alpha_\nu F(p_\nu)+F^\ast(c^\ast)-c^\ast c=J_{F,\boldsymbol{\alpha}}(\bm{p}).
\end{align}

Because $\alpha_\nu> 0(\nu=1,2,3\cdots,N)$ and the Bregman divergences have divergence properties, 
 $J_{F,\boldsymbol{\alpha}}(\bm{p})\geq 0$ and $J_{F,\boldsymbol{\alpha}}(\bm{p})=0\iff c=p_1=p_2=\cdots =p_N$ hold. 
Hence, the function defined by (\ref{eq_def_multi_Jensen}) has divergence properties.

Finally, we define the multivariate convex conjugate Jensen divergence.
\begin{defi}
\label{definition_of_conjugate_Jensen}
\begin{align}
\label{eq_definition_of_conjugate_Jensen}
J_{F^\ast,\boldsymbol{\alpha}}(\bm{p}^\ast)\eqdef\sum_\nu \alpha_\nu F^\ast(p_\nu^\ast)-F^\ast(\hat{c}^\ast)\\
\hat{c}^\ast\eqdef \sum_\nu \alpha_\nu p_\nu^\ast=\sum_\nu \alpha_\nu F'(p_\nu),
\end{align}
where $\bm{p}^\ast$ denotes a vector $(p_1^\ast,p_2^\ast,\cdots, p_N^\ast)=(F'(p_1),F'(p_2),\cdots, F'(p_N))$.
\end{defi}
We show this function has divergence properties later.

\begin{prop}
\label{prop_conj_Jensen_Bregman}
\begin{align}
J_{F^\ast,\boldsymbol{\alpha}}(\bm{p}^\ast)=\sum_\nu \alpha_\nu F^\ast(p_\nu^\ast)+F(\hat{c})-\hat{c}\hat{c}^\ast, 
\end{align}
where $\hat{c}^\ast\eqdef \sum_\nu \alpha_\nu p_\nu^\ast$.
\end{prop}
This equation is easily proved by using (\ref{convex conjugate}).

\begin{prop}
\label{prop_conj_Jensen_Bregman}
The multivariate convex conjugate Jensen divergence is equal to the sum of the Bregman divergences.
\begin{align}
J_{F^\ast,\boldsymbol{\alpha}}(\bm{p}^\ast)=\sum_\nu \alpha_\nu B_{F^\ast}(p_\nu^\ast, \hat{c}^\ast) = \sum_\nu \alpha_\nu B_F(\hat{c}, p_\nu)
\end{align}
\end{prop}
By using Proposition \ref{prop_reverse_Bregman}, the proof is the same as Proposition \ref{prop_Jensen_Bregman}.
Hence, the function defined by (\ref{eq_definition_of_conjugate_Jensen}) has divergence properties.

\begin{prop}
\label{prop_comp_Bregman}
\begin{align}
\sum_\nu \alpha_\nu B_F(c, p_\nu)=\sum_\nu \alpha_\nu B_F(\hat{c}, p_\nu)+B_F(c,\hat{c})
\end{align}
\end{prop}
\noindent\textbf{Proof.}
By using conjugate representation (\ref{Bregman_conjugate}) and $\sum_\nu \alpha_\nu=1$, we get
\begin{align}
\sum_\nu \alpha_\nu B_F(\hat{c}, p_\nu)=F(\hat{c})+\sum_\nu \alpha_\nu F^\ast(p_\nu^\ast)-\hat{c}\hat{c}^\ast.
\end{align}
In the same way, we get
\begin{align}
\sum_\nu \alpha_\nu B_F(c, p_\nu)=F(c)+\sum_\nu \alpha_\nu F^\ast(p_\nu^\ast)-c\hat{c}^\ast \\=\biggl(F(\hat{c})+\sum_\nu \alpha_\nu F^\ast(p_\nu^\ast)-\hat{c}\hat{c}^\ast\biggr)+\biggl(F(c)-F(\hat{c})-\hat{c}^\ast(c-\hat{c})\biggr)\\=\sum_\nu \alpha_\nu B_F(\hat{c}, p_\nu)
+B_F(c,\hat{c}).
\end{align}

\begin{theorem}(Basic sum decomposition theorem)
\label{th_base_decomp}
The sum of the 1-dimensional symmetric Bregman divergences is decomposed into the sum of three divergences.
\begin{align}
\sum_\nu\alpha_\nu B_{F,\mathrm{sym}}(p_\nu, c)=J_{F,\boldsymbol{\alpha}}(\bm{p})+J_{F^\ast,\boldsymbol{\alpha}}(\bm{p}^\ast)+B_F(c,\hat{c}) 
\end{align}
or
\begin{align}
\sum_\nu\alpha_\nu p_\nu^\ast(p_\nu-c)=J_{F,\boldsymbol{\alpha}}(\bm{p})+J_{F^\ast,\boldsymbol{\alpha}}(\bm{p}^\ast)+B_F(c,\hat{c}),
\end{align}
where $c\eqdef  \sum_\nu \alpha_\nu p_\nu$, $\hat{c}^\ast\eqdef\sum_\nu \alpha_\nu p_\nu^\ast$ and $\bm{p}^\ast$ denotes a vector $(p_1^\ast,p_2^\ast,\cdots, p_N^\ast)$.
\end{theorem}
\noindent\textbf{Proof.}
We prove the first equation.
From the definition of the 1-dimensional symmetric Bregman divergence (\ref{eq_def_1dim_sym_Bregman}) and Proposition \ref{prop_comp_Bregman}, we get
\begin{align}
\sum_\nu\alpha_\nu B_{F,\mathrm{sym}}(p_\nu, c)=\sum_\nu\alpha_\nu \bigl(B_F(p_\nu, c) +B_F(\hat{c},p_\nu)\bigr) + B_F(c,\hat{c}).
\end{align}
From Proposition \ref{prop_Jensen_Bregman} and \ref{prop_conj_Jensen_Bregman}, we get the first equation of Thoerem \ref{th_base_decomp} as follows.
\begin{align}
\sum_\nu\alpha_\nu B_{F,\mathrm{sym}}(p_\nu, c)&=\sum_\nu\alpha_\nu \bigl(B_F(p_\nu, c) +B_F(\hat{c},p_\nu)\bigr) + B_F(c,\hat{c}) \\ \nonumber
&=J_{F,\boldsymbol{\alpha}}(\bm{p})+J_{F^\ast,\boldsymbol{\alpha}}(\bm{p}^\ast)+B_F(c,\hat{c}) 
\end{align}
About the second equation, by using (\ref{eq_sym_Bregman_conjugate}) and $\sum_\nu \alpha_\nu(p_\nu-c)=0$, we get
\begin{align}
\sum_\nu\alpha_\nu B_{F,\mathrm{sym}}(p_\nu, c)=\sum_\nu\alpha_\nu (p_\nu^\ast-c^\ast)(p_\nu-c)=\sum_\nu\alpha_\nu p_\nu^\ast(p_\nu-c).
\end{align}
Hence, the result follows.

\section{Sum decomposition of the symmetric Bregman divergences}
\label{sec_sum_decomp}
\textbf{Notation of this section}
\begin{itemize}
\item Let $F:\mathrm{dom} F\rightarrow\mathbb{R}$ be differentiable and strictly convex functions and let $\mathrm{dom}F\subseteq\mathbb{R}$.
\item Let $i\in\{1,2,\cdots,M\}$ and $p,q,p_i\in\mathrm{dom}F$.
\item Let $P=\{p_1,p_2,\cdots,p_M\}$ and $Q=\{q_1,q_2,\cdots,q_M\}$.
\item Let $\alpha\in(0,1)$.
\end{itemize}
\subsection{Sum decomposition theorem 1}
In this subsection, we derive the sum decomposition theorem for the symmetric Bregman divergences.
\begin{defi}
\label{def_Bregman}
We define the Bregman and the symmetric Bregman divergences by using the 1-dimensional version defined in the section \ref{sec_base_sum_decomp}.
\begin{align}
B_F(P,Q)\eqdef \sum_i B_F(p_i,q_i)\\
\label{eq_def_sym_Bregman}
B_{F,\mathrm{sym}}(P,Q)\eqdef \sum_i B_{F,\mathrm{sym}}(p_i,q_i)
\end{align}
\end{defi}
For $N=2$, the multivariate Jensen and the multivariate convex conjugate Jensen divergences defined in section \ref{sec_base_sum_decomp} are
\begin{align}
J_{F,\alpha}(p,q)=\alpha F(p)+(1-\alpha) F(q)-F(c)\\
c\eqdef  \alpha p+(1-\alpha) q
\end{align}
and 
\begin{align}
J_{F^\ast,\alpha}(p^\ast,q^\ast)=\alpha F^\ast(p^\ast)+(1-\alpha) F^\ast(q^\ast)-F^\ast(c^\ast)\\
c^\ast\eqdef  \alpha p^\ast+(1-\alpha) q^\ast.
\end{align}

We define the Jensen and the convex conjugate Jensen divergences as follows.
\begin{defi}
\label{def_Jensen}
\begin{align}
\label{eq_def_Jensen}
J_{F,\alpha}(P,Q)\eqdef \sum_i J_{F,\alpha}(p_i,q_i)\\
\label{eq_def_conj_Jensen}
J_{F^\ast,\alpha}(P^\ast,Q^\ast)\eqdef\sum_i J_{F^\ast,\alpha}(p_i^\ast,q_i^\ast)
\end{align}
\end{defi}

\begin{prop}
The Jensen and the convex conjugate Jensen divergences can be represented as the weighted average of the Bregman divergences\cite{nielsen2011burbea}.
\begin{align}
J_{F,\alpha}(P,Q)=\alpha B_F(P,C)+(1-\alpha) B_F(Q,C)  \\
J_{F^\ast,\alpha}(P^\ast, Q^\ast)=\alpha B_{F^\ast}(P^\ast, \hat{C}^\ast)+ (1-\alpha)B_{F^\ast}(Q^\ast, \hat{C}^\ast)\\
=\alpha B_F(\hat{C},P)+ (1-\alpha)B_F(\hat{C},Q),
\end{align}
where $C$ denotes $\{c_i\eqdef \alpha p_i+ (1-\alpha)q_i\}$ and $\hat{C}$ denotes $\{\hat{c}_i^\ast\eqdef \alpha p_i^\ast+ (1-\alpha) q_i^\ast\}$.
\end{prop}
\noindent\textbf{Proof.}
From Proposition \ref{prop_Jensen_Bregman}, we get $J_{F,\alpha}(p_i,q_i)=\alpha B_F(p_i,c_i)+(1-\alpha) B_F(q_i,c_i)$. 
Taking the sum of the subscript $i$, the first equation follows. We can prove the second equation in the same way by applying Proposition \ref{prop_conj_Jensen_Bregman}.

\begin{theorem}(Sum decomposition theorem 1)
\label{th_decomp}
The symmetric Bregman divergence is decomposed into the sum of three divergences.
\begin{align}
\alpha(1-\alpha)B_{F,\mathrm{sym}}(P,Q)=J_{F,\alpha}(P,Q)+J_{F^\ast,\alpha}(P^\ast,Q^\ast)+B_F(C,\hat{C}) ,
\end{align}
where $C$ denotes $\{c_i\eqdef \alpha p_i+ (1-\alpha) q_i\}$ and $\hat{C}$ denotes $\{c_i^\ast\eqdef \alpha p_i^\ast+(1-\alpha) q_i^\ast\}$.
\end{theorem}
\noindent\textbf{Proof.}
From the second equation of Theorem \ref{th_base_decomp}, we get
\begin{align}
\label{eq_th_decomp_1}
\alpha p_i^\ast(p_i-c_i)+(1-\alpha) q_i^\ast(q_i-c_i)=J_{F,\alpha}(p_i,q_i)+J_{F^\ast,\alpha}(p_i^\ast,q_i^\ast)+B_F(c_i,\hat{c_i}).
\end{align}
By using the definition of $c_i$, we obtain 
\begin{align}
p_i-c_i=(1-\alpha)(p_i-q_i) \\
q_i-c_i=-\alpha(p_i-q_i).
\end{align}
Substituting these equations to (\ref{eq_th_decomp_1}) and using (\ref{eq_sym_Bregman_conjugate}), we obtain
\begin{align}
\alpha(1-\alpha)B_{F,\mathrm{sym}}(p_i,q_i)=J_{F,\alpha}(p_i,q_i)+J_{F^\ast,\alpha}(p_i^\ast,q_i^\ast)+B_F(c_i,\hat{c_i}).
\end{align}
Taking the sum of the subscript $i$, we get the result.

\begin{corollary}
\label{cor_decomp_inequality}
The following inequalities hold.
\begin{align}
\alpha(1-\alpha)B_{F,\mathrm{sym}}(P,Q)\geq J_{F,\alpha}(P,Q) \\
\alpha(1-\alpha)B_{F,\mathrm{sym}}(P,Q)\geq J_{F^\ast,\alpha}(P^\ast,Q^\ast) \\
\alpha(1-\alpha)B_{F,\mathrm{sym}}(P,Q)\geq B_F(C,\hat{C})
\end{align}
\end{corollary}
\noindent\textbf{Proof.}
Because $J_{F,\alpha}(P,Q)$,  $J_{F^\ast,\alpha}(P^\ast,Q^\ast)$ and $B_F(C,\hat{C})$ are non-negative, the result follows.
The first inequality has shown in \cite{nishiyama2018generalized}.

\subsection{Example}
We show an example for $F(x)=x\log x$.
In this case, the following equations hold.
\begin{itemize}
\item $F^\ast(x^\ast)=\exp(x^\ast-1)$.
\item $x^\ast=F'(x)=\log x + 1$.
\item $c=\alpha p + (1-\alpha)q$.
\item $\hat{c}^\ast=\log(p^\alpha q^{1-\alpha}) + 1$.
\item $B_F(P,Q)=\mathrm{KL}(P\| Q)$.\\
$\mathrm{KL}(P,Q)\eqdef -\sum_i p_i + \sum_i q_i + \sum_i p_i \log \bigl(\frac{p_i}{qi}\bigr)$: Kullback-Leibler divergence.
\item $B_{F,\mathrm{sym}}(P,Q)=\mathrm{J}(P,Q)$.\\
$\mathrm{J}(P,Q)\eqdef \mathrm{KL}(P\| Q)+\mathrm{KL}(Q\| P)$: Jeffrey's J-divergence.
\item $J_{F,\alpha}(P,Q)=\alpha B_F(P,C)+(1-\alpha)B_F(Q,C)=\mathrm{JS}_\alpha(P\|Q)$.\\
 $\mathrm{JS}_\alpha(P\|Q)\eqdef \alpha \mathrm{KL}(P\| C) + (1-\alpha) \mathrm{KL}(Q\| C)$: skew Jensen-Shannon divergence.
\item  $J_{F^\ast,\alpha}(P^\ast,Q^\ast)=\sum_i \bigl(\alpha F^\ast(p_i^\ast) +(1-\alpha) F^\ast(q_i^\ast) - F^\ast(\hat{c_i}^\ast)\bigr)=\alpha(1-\alpha)D_\alpha(P\|Q)$.\\
$D_\alpha(P\|Q)\eqdef \frac{1}{\alpha(\alpha - 1)} \bigl(\sum_i p_i^\alpha q_i^{1-\alpha} - \alpha \sum_i p_i - (1-\alpha) \sum_i q_i\bigr)$: Amari's alpha-divergence\cite{cichocki2010families}.
\item $B_F(C,\hat{C})=\mathrm{KL}(\alpha P+(1-\alpha) Q\| P^\alpha Q^{1-\alpha})\eqdef\sum_i \mathrm{KL}(\alpha p_i+(1-\alpha) q_i\| p_i^\alpha q_i^{1-\alpha})$.
\end{itemize}

By using these results and applying Theorem \ref{th_decomp}, we get the following decomposition equation.
\begin{corollary}
\begin{align}
\alpha(1-\alpha) \mathrm{J}(P,Q)=\mathrm{JS}_\alpha(P\|Q)+\alpha(1-\alpha) D_\alpha(P\|Q) +\mathrm{KL}(\alpha P+(1-\alpha) Q\| P^\alpha Q^{1-\alpha})
\end{align}
When $\alpha=\frac{1}{2}$, 
\begin{align}
\frac{1}{4}\mathrm{J}(P,Q)=\mathrm{JS}(P,Q)+\mathrm{H}^2(P,Q) + \mathrm{KL}\bigl(\frac{1}{2}(P+Q)\| \sqrt{PQ}\bigr),
\end{align}
where $\mathrm{JS}(P,Q)\eqdef \mathrm{JS}_{\frac{1}{2}}(P\|Q)$ is the Jensen-Shannon divergence and $\mathrm{H}^2(P,Q)\eqdef \frac{1}{2}\sum_i(\sqrt{p_i}-\sqrt{q_i})^2$ is the squared Hellinger distance.
\end{corollary}
From this corollary and Corollary \ref{cor_decomp_inequality}, we get
\begin{align}
\label{eq_Lin_inequality}
\frac{1}{4}\mathrm{J}(P,Q)\geq \mathrm{JS}(P,Q). 
\end{align}
This the Lin's inequality\cite{lin1991divergence}.

\section{Sum decomposition of divergences including f-divergences explicitly}
\textbf{Notation of this section}
\begin{itemize}
\item Let $F:\mathrm{dom} F\rightarrow\mathbb{R}$ be differentiable and strictly convex functions and let $\mathrm{dom}F\subseteq\mathbb{R_{++}}$.
\item Let $i\in\{1,2,\cdots,M\}$ and $p_i , q_i\in \mathrm{dom}F$.
\item Let $P=\{p_1,p_2,\cdots,p_M\}$ and $Q=\{q_1,q_2,\cdots,q_M\}$.
\item Let $S\eqdef \sum_i q_i  > 0$ and let $\{p_i\}$ satisfy $\sum_i p_i = \sum_i q_i$.
\end{itemize}

\subsection{Sum decomposition theorem 2}
In this subsection, we derive the sum decomposition theorem which explicitly includes the $f$-divergences.

By replacing $\nu\rightarrow i$, $N\rightarrow M$, $p_\nu\rightarrow \frac{p_i}{q_i}$ and putting $\alpha_i=\frac{q_i}{S}$, we define two functions as follows.
\begin{defi}
\label{def_Df}
\begin{align}
\label{eq_def_Df}
D_F(P,Q)\eqdef J_{F,\boldsymbol{\alpha}}\bigl(\frac{p_1}{q_1},\frac{p_2}{q_2},\cdots ,\frac{p_M}{q_M}\bigr)=\frac{1}{S}\sum_i q_i F\bigl(\frac{p_i}{q_i}\bigr) -F(1) \\
\label{eq_def_conj_Df}
\hat{D}_F(P,Q)\eqdef J_{F^\ast,\boldsymbol{\alpha}}\biggl({\bigl(\frac{p_1}{q_1}\bigr)}^\ast,{\bigl(\frac{p_2}{q_2}\bigr)}^\ast,\cdots ,{\bigl(\frac{p_M}{q_M}\bigr)}^\ast\biggr)=\frac{1}{S}\sum_i q_i F^\ast\bigl(F'\bigl(\frac{p_i}{q_i}\bigr)\bigr) -F^\ast(\hat{c}),
\end{align}
where ${\bigl(\frac{p_i}{q_i}\bigr)}^\ast$ denotes $F'\bigl(\frac{p_i}{q_i}\bigr)$ and $\hat{c}\eqdef \frac{1}{S}\sum_i q_i F'\bigl(\frac{p_i}{q_i}\bigr)$.
\end{defi}

\begin{prop}
 $D_F(P,Q)$ and $\hat{D}_F(P,Q)$ are divergence functions.
\end{prop}
\noindent\textbf{Proof.}
From the definition, $D_F(P,Q)$ and $\hat{D}_F(P,Q)$ are non-negative and $D_F(P,Q)=0$ or  $\hat{D}_F(P,Q)=0$ holds if and only if $\frac{p_1}{q_1}=\frac{p_2}{q_2}=\cdots=\frac{p_M}{q_M}$.
By the assumption of this section $\sum_i p_i = \sum_i q_i$,  $D_F(P,Q)=0$ or  $\hat{D}_F(P,Q)=0$ holds  if and only if $P=Q$.
Hence, $D_F(P,Q)$ and $\hat{D}_F(P,Q)$ are divergence functions.

\begin{theorem}(Sum decomposition theorem 2)
\label{th_decomp_for_f}
For $D_F(P,Q)$ and $\hat{D}_F(P,Q)$, the following equation holds.
\begin{align}
\frac{1}{S}\sum_i F'\bigl(\frac{p_i}{q_i}\bigr)(p_i-q_i)=D_F(P,Q)+\hat{D}_F(P,Q)+B_F(1,\hat{c}),
\end{align}
where $\hat{c}^\ast\eqdef \frac{1}{S}\sum_i q_i F'\bigl(\frac{p_i}{q_i}\bigr)$.
\end{theorem}
\noindent\textbf{Proof.}

By replacing $\nu\rightarrow i$, $N\rightarrow M$, $p_\nu\rightarrow \frac{p_i}{q_i}$ and putting $\alpha_i=\frac{q_i}{S}$, $c\eqdef \sum_\nu \alpha_\nu p_\nu$ and $\hat{c}^\ast \eqdef \sum_\nu \alpha_\nu p_\nu^\ast $ in Theorem \ref{th_base_decomp} are equal to 
$1$ and $\frac{1}{S}\sum_i q_i F'\bigl(\frac{p_i}{q_i}\bigr)$ respectively.
By using the second equation of Theorem \ref{th_base_decomp} and the definition of $D_F(P,Q)$ and $\hat{D}_F(P,Q)$, we get 
\begin{align}
\label{eq_th_decomp_2}
\sum_i \frac{q_i}{S} {\bigl(\frac{p_i}{q_i}\bigr)}^\ast(\frac{p_i}{q_i}-1)=D_F(P,Q)+\hat{D}_F(P,Q)+B_F(1,\hat{c}).
\end{align}
Calculating the LHS of this equation and using $x^\ast=F'(x)$, the result follows.
When $S=1$ and $F(1)=0$, $D_F(P,Q)$ is consistent with the $f$-divergence.
\begin{corollary}
\label{cor_decomp_inequality_for_f}
The following inequalities hold.
\begin{align}
\frac{1}{S}\sum_i F'\bigl(\frac{p_i}{q_i}\bigr)(p_i-q_i)\geq D_F(P,Q) \\
\frac{1}{S}\sum_i F'\bigl(\frac{p_i}{q_i}\bigr)(p_i-q_i)\geq \hat{D}_F(P,Q) \\
\frac{1}{S}\sum_i F'\bigl(\frac{p_i}{q_i}\bigr)(p_i-q_i)\geq B_F(1,\hat{c})
\end{align}
\end{corollary}
\noindent\textbf{Proof.}
Because $D_F(P,Q) $,  $\hat{D}_F(P,Q)$ and $B_F(1,\hat{c})$ are non-negative, the result follows.
The first inequality in corollary \ref{cor_decomp_inequality_for_f} is well-known inequality of $f$-divergences\cite{dragomir2000inequalities}.
\subsection{Example}
We show an example for $F(x)=-\log x$ and $S=1$.
In this case, the following equations hold.
\begin{itemize}
\item $F^\ast(x^\ast)=-\log(-x^\ast)-1$.
\item $x^\ast=F'(x)=-\frac{1}{x}$.
\item $\hat{c}^\ast=-\sum_i \frac{q_i^2}{p_i}=-(\chi_N^2(P\|Q)+1)$.\\
$\chi_N^2(P\|Q)\eqdef \sum_i \frac{(p_i-q_i)^2}{p_i}$: Neyman chi-square divergence.

\item $\sum_i F'\bigl(\frac{p_i}{q_i}\bigr)(p_i-q_i)=\chi_N^2(P\|Q)$.
\item $D_F(P,Q)=\mathrm{KL}(Q\| P)$.
\item $\hat{D}_F(P,Q)=\log(\chi_N^2(P\|Q)+1)-\mathrm{KL}(Q\| P)$.
\item $B_F(1,\hat{c})=\mathrm{IS}\bigl(1\|\frac{1}{\chi_N^2(P\|Q)+1}\bigr)$.\\
$\mathrm{IS}(p,q)\eqdef \frac{p}{q}-\log\frac{p}{q} -1$: Itakura-Saito divergence.
\end{itemize}
By using these results and applying Theorem \ref{th_decomp_for_f}, we get the following decomposition equation.
\begin{corollary}
\begin{align}
\chi_N^2(P\|Q)=\mathrm{KL}(Q\| P)+\mathrm{IS}\biggl(1\|\frac{1}{\chi_N^2(P\|Q)+1}\biggr)+\biggl(\log\bigl(\chi_N^2(P\|Q)+1\bigr)-\mathrm{KL}(Q\| P)\biggr)
\end{align}
\end{corollary}
Because the last term of RHS is a divergence function, we also get the following inequality.
\begin{corollary}
\begin{align}
\chi_N^2(P\|Q)\geq \exp(\mathrm{KL}(Q\| P))-1
\end{align}
\end{corollary}

\section{Conclusion}
We have shown the sum decomposition of divergence into three divergences.

First, we have shown a basic sum decomposition theorem that the sum of the 1-dimensional symmetric Bregman divergences is decomposed into the multivariate Jensen, the convex conjugate multivariate Jensen and the 1-dimensional Bregman divergences.

Next, by using this result, we have shown that the symmetric Bregman divergence is decomposed into the sum of three divergences which are the Jensen, the convex conjugate Jensen and the Bregman divergences.

Finally, by applying the basic sum decomposition theorem, we have derived the equation of the sum decomposition which explicitly includes $f$-divergences.

It is expected that these results make the relationship between many divergences clearer.
\bibliography{reference_v1}
\end{document}